\documentclass[]{spie}  
\usepackage[]{graphicx}
\usepackage[]{amsmath}
\usepackage[]{amstext}

\newcommand{\degree}{\ensuremath{^\circ}}

\usepackage{color}
\definecolor{blue}{rgb}{0.1,0.1,0.6}
\definecolor{orange}{rgb}{0.74,.35,0.099}
\definecolor{pale}{rgb}{0.90,0.90,0.95}
\definecolor{red}{rgb}{1.0,0.0,0.0}

\title{Gemini Planet Imager Observational Calibrations II: \\ Detector Performance and Calibration}
\author{Patrick Ingraham\supit{ab}, Marshall D. Perrin\supit{c}, Naru Sadakuni\supit{d}, Jean-Baptiste Ruffio\supit{ef}, Jerome Maire\supit{g}, Jeff Chilcote\supit{h}, James Larkin\supit{h}, Franck Marchis\supit{e}, Raphael Galicher\supit{i}, Jason Weiss\supit{h}
\skiplinehalf
\supit{a}Kavli Institute for Particle Astrophysics and Cosmology, Stanford University, Stanford, CA 94305, USA \\
\supit{b}Department de Physique, Universit\'{e} de Montr{\'e}al, Montr\'eal QC H3C 3J7, Canada \\
\supit{c}Space Telescope Science Institute, 3700 San Martin Drive, Baltimore MD 21218 USA \\
\supit{d}Gemini Observatory, Casilla 603, La Serena, Chile \\
\supit{e}SETI Institute, Carl Sagan Center, 189 Bernardo Avenue, Mountain View, CA 94043, USA\\
\supit{f}Institute Superieur de l'Aeronautique et de l'Espace, Toulouse, France \\  
\supit{g}Dunlap Institute for Astrophysics, University of Toronto, 50 St. George St, Toronto ON M5S 3H4, Canada \\
\supit{h}Univeristy of California Los Angeles, Los Angeles California, USA \\
\supit{i}LESIA, CNRS, Observatoire de Paris, Univ. Paris Diderot, UPMC, 5 place Jules Janssen, 92190 Meudon, France
}
\authorinfo{Further author information: contact Patrick Ingraham E-mail: patricki@stanford.edu }

\begin{document}
\maketitle 

\begin{abstract}
The Gemini Planet Imager is a newly commissioned facility instrument designed to measure the near-infrared spectra of young extrasolar planets in the solar neighborhood and obtain imaging polarimetry of circumstellar disks. GPI's science instrument is an integral field spectrograph that utilizes a HAWAII-2RG detector with a SIDECAR ASIC readout system. This paper describes the detector characterization and calibrations performed by the GPI Data Reduction Pipeline to compensate for effects including bad/hot/cold pixels, persistence, non-linearity, vibration induced microphonics and correlated read noise.
\end{abstract}



\section{Introduction}
\label{sec:intro}

The Gemini Planet Imager\cite{Macintosh14a}\cite{Macintoshthis} is a newly commissioned instrument at Gemini South Observatory that combines an extreme adaptive optics system, an advanced coronagraph, and an integral field spectrograph (IFS) to measure the near-infrared spectra of young extrasolar planets orbiting stars in the stellar neighborhood ($<$100 pc), and conduct related observations such as imaging polarimetry of circumstellar disks. The science detector in its infrared IFS\cite{Chilcote12a}\cite{Larkinthis} is a 2048$\times$2048 pixel HgCdTe HAWAII-2RG (H2RG) with an 18 $\mu$m pixel pitch\cite{Beletic08a}.  The readout electronics feature an on-chip SIDECAR ASIC that enables cryogenic operations with low power and low read noise\cite{Loose07a}.

In this paper, we discuss the properties and noise characteristics of the GPI H2RG detector, including its readnoise, dark current, and bad pixels. We report our preliminary efforts to characterize and implement corrections for both the detector persistence and non-linearity. We present a novel method to remove the correlated noise from dark frames and in GPI science frames, where the correlated noise becomes significantly more challenging to extract due to the presence of the complicated intensity pattern resulting from the thousands of lenslet spectra. We also present a method to measure and remove microphonics from individual science images via Fourier filtering. All of these corrections are performed using the GPI Data Reduction Pipeline\cite{perrinthis} (DRP), an open source software package written in IDL to reduce GPI data from raw files into science-ready calibrated datacubes. The implementation and algorithms for the detector calibrations in the GPI DRP are discussed here, with additional details available in the pipeline documentation online. 

\section{Basic properties and operation}

For the specific detector used in GPI, the HgCdTe semiconductor band gap is tuned to extend the wavelength coverage over 0.9-2.5 $\mu$m, with a measured cutoff wavelength of 2.56 $\mu$m\footnote{As measured on a process evaluation chip, based on data provided by the vendor}.  The quantum efficiency ranges from 79\%-92\% over that wavelength range, as measured by Teledyne. The CdZnTe detector substrate (used to hold the HgCdTe during its deposition) is removed to boost the quantum efficiency (QE) at short wavelengths and greatly reduce cosmic ray count rates. The GPI detector has excellent cosmetic performance with $<$0.5\% inoperable pixels, no large clusters of contiguous bad pixels, low dark current and interpixel capacitance.

GPI operates the detector using 32 readout channels in the up-the-ramp (UTR) mode for all observations, which rejects (\textit{kTC}) noise and reduces other noise sources. Pixels are clocked out at 100 kHz, resulting in a readout time of 1.45479 s per full frame readout; all exposure times are quantized to be multiples of this, so the possible exposure times are $\sim$1.45 s, 2.91 s, 4.35 s, and so on. Overhead from resetting the array before an exposure is 1 readout time, plus typically half of a readout time waiting for the start of a readout time boundary before initiating the reset, plus 1 readout time for the initial read of the UTR sequence,  
for a total intrinsic overhead of 2.5 readout times ($\sim$3.75 s) for UTR exposure. To this is added software overheads for commanding, setup and file writing, resulting in a total overhead of order 15 s per exposure written to disk. For greater efficiency with short exposures, multiple readouts can be coadded together before output, which incurs only the reset+read overhead of 3 s per coadd. The hardware supports subarray readouts but at a significant overhead penalty, so this capability has not yet been commissioned. Correlated double sampling (CDS) and Multiple Correlated Double Sampling (Fowler Sampling) are also supported but are not generally used as the UTR mode delivers higher S/N per unit time.  

The entire GPI IFS cryostat is cooled by two Sunpower Stirling cycle cryocoolers to bring the detector to a temperature of $\sim$80 K.
Vibrations induced by the cryocoolers can at times couple into the detector to produce measurable microphonics noise. Several other well-known correlated noise patterns are also known to exist, such as the $1/f$ noise and the bias offsets between the individual channels\cite{Loose10a} \cite{Grogin10a}. These are discussed further below.

GPI FITS files are written out as standard Gemini Multi-Extension FITS files. The primary header contains keywords recording the overall GPI and observatory status. An image extension ``SCI'' contains the science image as a 2048$\times$2048 floating point array, and an image extension ``DQ'' provides data quality flags for each pixel\footnote{See http://docs.planetimager.org/pipeline/ifs/dataformats.html for details of the DQ logical flags byte format.}.  The IFS instrument software performs in real time the UTR slope fitting process using a memory-efficient algorithm based on precomputed weights for each read, and in normal operation only the resulting slope value is written for each pixel. This is written out scaled by the total exposure time such that the units of output files are total counts, rather than counts/second, but we emphasize that the values output are based on the entire series of detector reads. There is also a write-all mode that writes every read of the  sequence to disk. This is particularly useful when a high-frequency (time) sampling is desired, a capability we have used to measure detector properties such as non-linearity and persistence.

HAWAII-2RG detectors have a 4-pixel-wide ring of non-photosensitive reference pixels around the outer edge of the field of view. The reference pixels at the top and bottom of the detector\footnote{I.e. the reference pixels in the first four rows and the last four rows read out in each channel.} are used to measure and remove the bias offsets between the readout channels. This is done by the IFS instrument software in real time as part of the slope fitting process just described, so the FITS files written out have already had vertical reference pixel correction applied. The reference pixels in the first four and last four columns are \textit{not} used for horizontal stripe removal, since that correction is better made by the destriping steps discussed below (\S \ref{sec:destriping}).

\section{Darks}

\begin{figure}[ht]
\begin{center}
\includegraphics[height=8cm, trim=0cm 0cm 0cm 0cm]{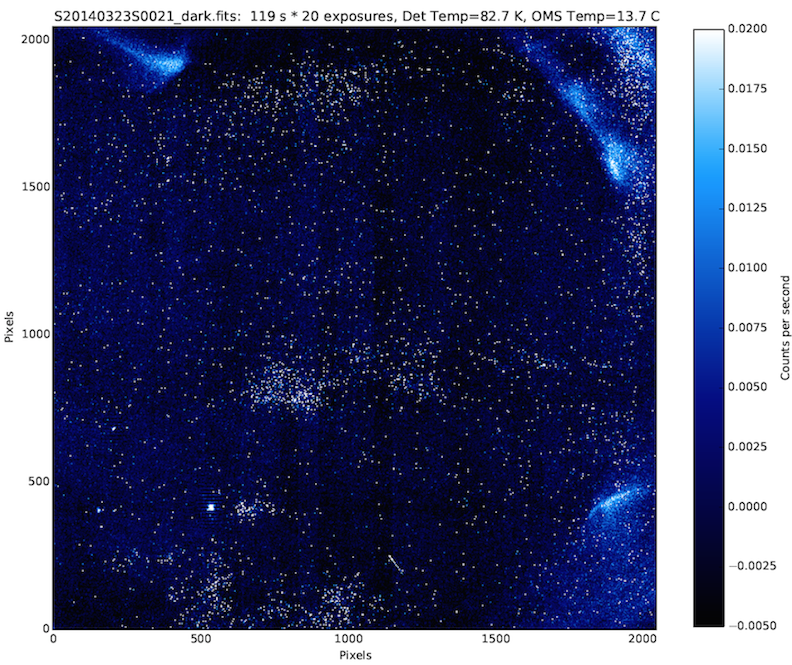} 
\end{center}
\caption[example]{ \label{fig:dark_frame} GPI dark image. Very long exposure, high SNR dark frames demonstrate a very low-level light leak at a rate of $\sim$0.02 counts/sec/pixel near three of the corners. This has zero effect on science images which are generally less than 60-120 seconds. Hot pixels are scattered across the array. The dark count rate for normal pixels is negligible and measured to be $\sim$0.006 counts/sec/pixel.}
\end{figure} 

Like all infrared (IR) detectors, or any detector based on the photoelectric effect, the HAWAII-2RG detector in GPI has nonzero dark current. For GPI, dark files are used to subtract dark current, the detector bias, and other background (e.g. light leaks) from science and calibration observations. Dark frames are taken by setting the IFS's Lyot stop to ``Blank'' (an opaque mask) and taking multiple exposures which are combined to increase signal to noise ratio\footnote{In practice it is also important to not schedule dark observations immediately after brightly illuminated observations that will cause persistence, for instance the very bright GCAL flat field lamps.}. Typical pixels have dark current much less than 1 count per second per pixel, which is negligible for essentially all GPI observations. However, some ``hot'' pixels have higher dark rates and necessitate dark subtraction.  

The ``Combine 2D Dark Images'' Data Reduction Pipeline primitive\footnote{We assume the reader is familiar with the basic operation of the GPI pipeline and key concepts such as reduction recipes and primitives. For description of these see \cite{perrinthis}} merges the individual darks into a master dark using a 3-sigma-clipped resistant mean to remove any possible cosmic ray events. While dark and hot pixel count rates are for the most part linear, because of the possible presence of various nonlinear detector effects (saturated hot pixels, reset anomaly, etc), it is generally considered better to use a reference dark that is similar in integration time to the science frame.  The standard set of facility darks taken by Gemini includes exposure times increasing by steps of 2-3$\times$ from 1.45 s to 120 s, obtained with multiple coadds and exposures to ensure the S/N surpasses the levels obtained during science observations. These allow dark subtraction of arbitrary exposure times with scaling factors always less than 2. The differential measurement techniques used in GPI (e.g. PSF subtraction by angular differential imaging) typically will automatically reject any residual hot pixels, but careful dark subtraction is beneficial for measurements that occur prior to PSF subtraction (such as localization of satellite spots) and for observations that do not use PSF subtraction such as solar system resolved bodies.

Periodic dark observations demonstrate that the detector's complement of hot pixels evolves slightly but noticeably on approximately week to week timescales as lattice defects within the crystal migrate stochastically. Using darks that differ in time by more than a week or two from science data results in visibly more residual hot pixels in the final datacubes. We have recommended weekly cadence for obtaining facility dark calibrations at Gemini.

The detector readout exhibits a nonzero bias, even after subtraction of the initial read during the UTR slope fit.  This is dominated by the ``reset anomaly'' effect settling in the first readout or two of the UTR series, which leads to a significant slope in the background across the first $\sim$ 100 rows of the image and an approximately uniform bias level elsewhere. This effect is very repeatable and is well subtracted by exposure-time-matched dark frames. Its apparent magnitude scales with $1/\sqrt{N_{reads}}$ so it has the greatest impact for short exposure times. The destriping process discussed below will remove any residual bias offset remaining after dark subtraction.

The GPI IFS is well baffled but exhibits a very low level light leak visible in long duration dark exposures as a diffuse pattern in three corners of the detector (see Figure \ref{fig:dark_frame}). This is believed to be scattered light leaking around a small gap in the detector housing between the detector itself and the field flattening lens. Its peak magnitude is about 0.02 counts/sec/pixel, comparable to the intrinsic dark current of typical pixels.   While this is visible on high signal-to-noise ratio (SNR) long exposure darks, in practice it has negligible impact on science.

\section{Correlated noise removal}
\label{sec:destriping}

\begin{figure}[ht]
\begin{center}
\begin{tabular}{cc}
\includegraphics[height=7cm, trim=0cm 0cm 0cm 0cm]{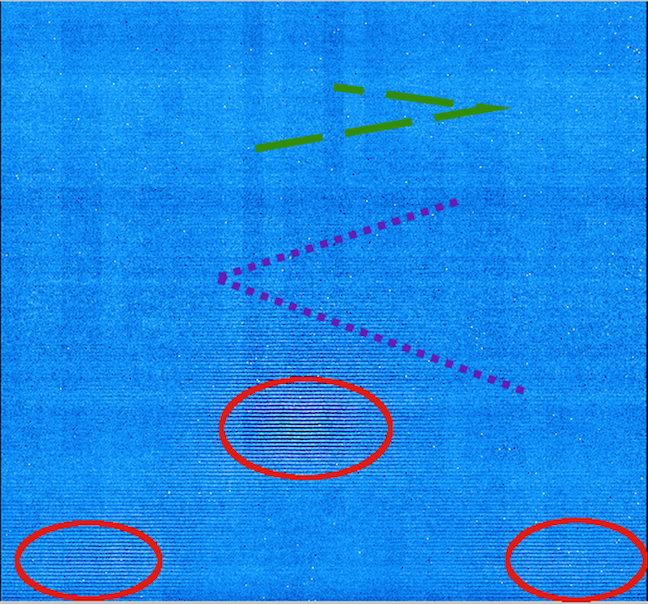} & \includegraphics[height=7cm, trim=0cm 0cm 0cm 0cm]{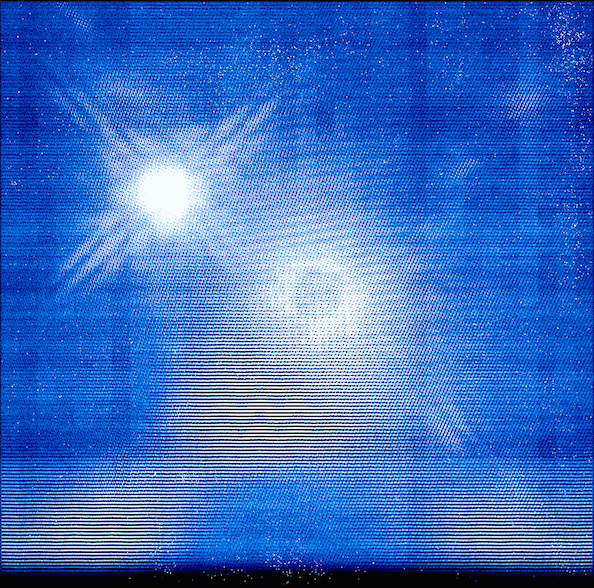}
\end{tabular}
\end{center}
\caption[example]{ \label{fig:correlated_noise} \textbf{Left:} A 1.5 second GPI dark frame shows the three independent sources of correlated noise. The red circles indicate sections of the detector heavily affected by the microphonic noise. The purple dashed lines indicate regions of horizontal detector striping and the green dashed lines indicate example of the vertical striping. \textbf{Right:} 10 coadds of a 1.5 second short exposure coronagraphic image exhibits similar features, but are much harder to identify. This image is heavily stretched  to make them visible.}
\end{figure}

Beyond the Gaussian read noise, the GPI detector is subject to three known  independent sources of correlated noise. These are most easily identified in a 1.5 second dark frame (Figure \ref{fig:correlated_noise} left), but are more difficult to observe in a longer closed-loop science image (Figure \ref{fig:correlated_noise} right). The three components are vertical striping, horizontal striping, and microphonics noise. The GPI pipeline includes primitives that derive noise models based on non-illuminated pixels and use them to attenuate all three sources of noise in both dark and science frames. The estimation method is dependent upon the amount of flux in the image. For a dark frame, no flux is present, and therefore the pipeline primitive, ``Destripe for Darks Only'' utilizes the entire image to derives a high-fidelity noise model. For the more complicated case of science exposures which have complex patterns of microspectra or polarization spots over much of the detector, the illuminated pixels must be masked out such that only the un-illuminated pixels are used to derive the correlated noise model. This is accomplished using the pipeline primitive, ``Destripe science image.''  With the exception of the masking, the algorithms to determine the noise model are similar between the dark and science image cases.

The vertical striping, indicated by the green-dashed lines in Figure \ref{fig:correlated_noise} (left) are a result of residual bias offsets in each of the 32 individual detector channels that are not corrected by the IFS server software using the reference pixels. Such residuals typically occur when the bias level of a given channel changes non-linearly during an exposure so it is not well fit using only the top and bottom reference pixel rows. The noise model for this component is derived by taking the median value of the un-illuminated pixels but future versions will employ low-order polynomial fitting.

The horizontal striping, indicated by the purple dotted lines in Figure \ref{fig:correlated_noise} (left), is due to the well known issue of $1/f$ noise in the readout electronics, primarily the SIDECAR ASIC \cite{Loose10a}\cite{Grogin10a}. Although the striping appears to be broad stripes across the entire detector, in fact the stripes are repeated in every detector channel alternating in direction due to the H2RG readout pattern. This repetition across channels is key to deriving the noise model. The algorithm folds then stacks the 32 channels, then performs a median of the stack to derive the noise model. Any missing pixels are modeled as the median value of the line of pixels. The user is able to set a threshold to ensure that if too many missing pixels are present, the destriping routine is aborted. 

\begin{figure}[ht]
\begin{center}
\begin{tabular}{cc}
\includegraphics[height=7cm, trim=0cm 0cm 0cm 0cm]{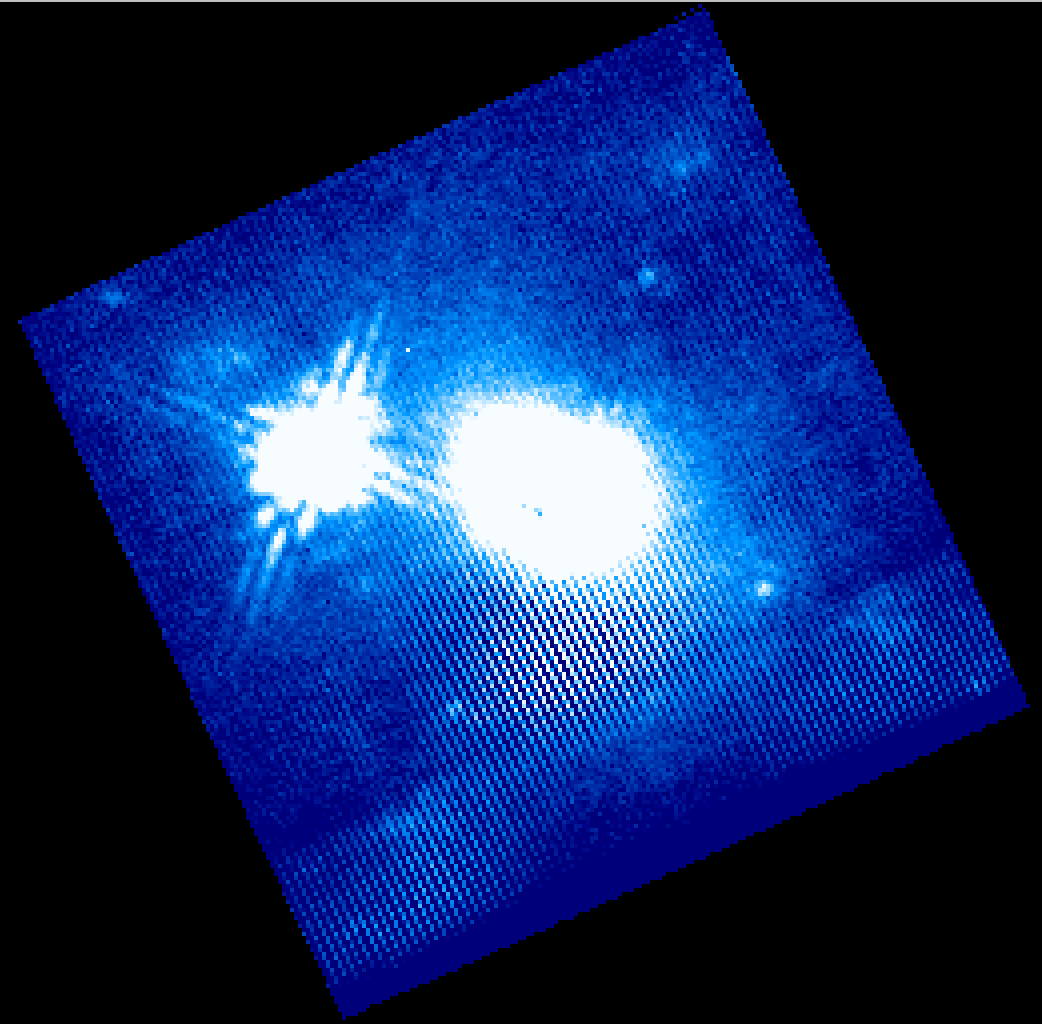} & \includegraphics[height=7cm, trim=0cm 0cm 0cm 0cm]{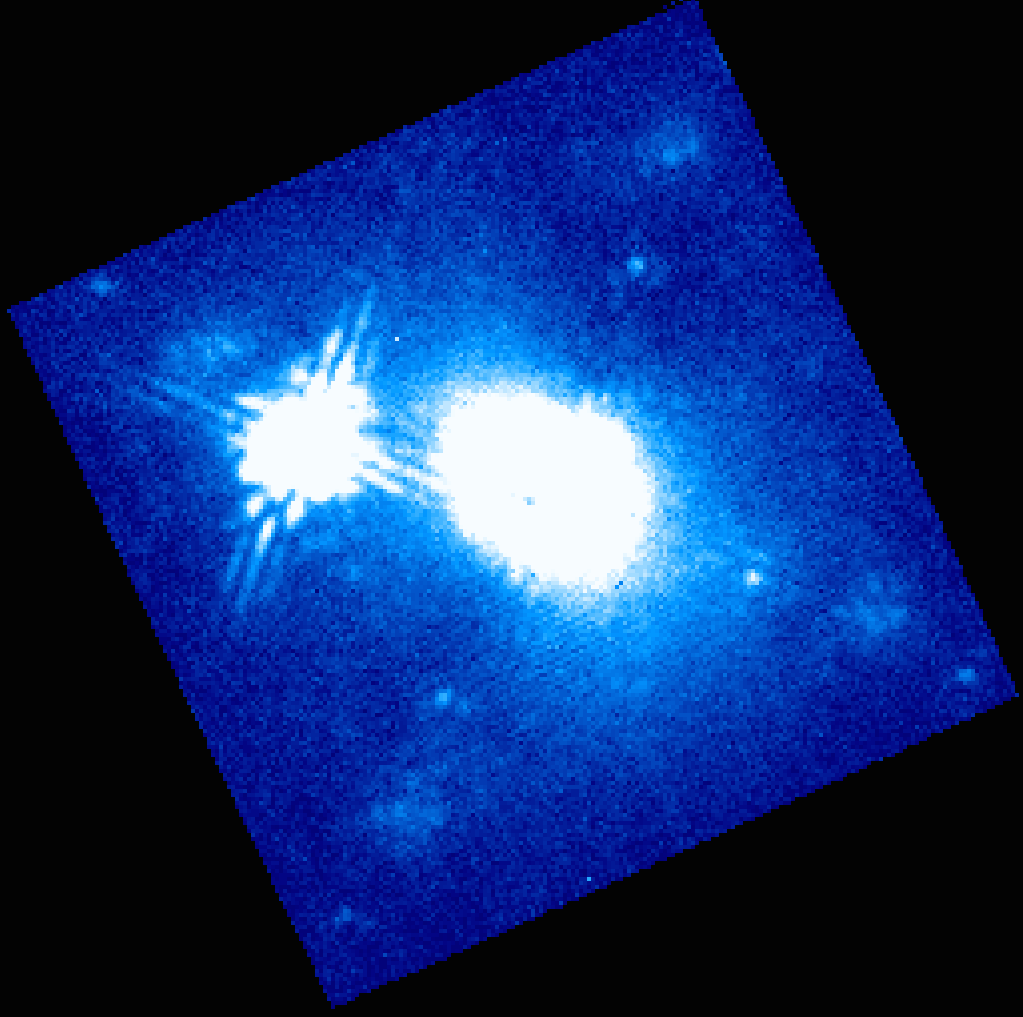}
\end{tabular}
\end{center}
\caption[example]{ \label{fig:before_after_destripe} \textbf{Left:} A quicklook reduced cube consisting of 10 coadds of a 1.5 second image which has heavily stretched to show the result of the correlated noise in reduced science images. \textbf{Right:} The same cube after running the destriping primitive (same image stretch). }
\end{figure}

The third source of noise is microphonics noise, indicated by the solid red circles in Figure \ref{fig:correlated_noise} (left), that results from vibrations from the cryo-coolers\cite{Chilcote12a}. GPI has two identical Sunpower cryocoolers which oscillate at $\sim$60 Hz. During integration and test, science verification, and the first two instrument commissioning observing runs, the two motors would beat in and out of phase and the pattern was observed to oscillate in intensity. After January 2014 a controller upgrade has synchronized the coolers to run always at 180\degree  out of phase, which has significantly reduced the microphonics noise but not eliminated it. The microphonics noise occurs in specific regions of the detector and with characteristic spatial frequencies that facilitate its filtering in Fourier space. 
Figure \ref{fig:before_after_destripe} demonstrates the efficiency of our algorithms in removing this pattern noise. 
However, if there is a large amount of structure in the image, such as for flat fields or arc-lamp images that have microspectra filling the entire detector, those can contribute power on the same spatial frequencies. Then
the microphonics do not occupy a unique frequency/phase space and cannot be removed cleanly; however for bright flat or arc lamp exposures the S/N is generally sufficiently high that microphonics noise can be neglected. 

Since the impact of the above correlated noise terms decreases as $1/\sqrt{N_{reads}}$, for moderately long exposure images ($>$30 seconds) the striping averages down to levels that often make destriping unnecessary. However, for short exposure images the destriping offers a significant reduction in the noise.

\section{Bad pixels}

The H2RG detector housed in the GPI IFS has approximately 99.5\% of its pixels having a normal response. This does not include the 4-pixel wide rim of reference pixels around the outer edge of the detector which make up 0.78\% of the total pixels, but are non-photosensitive by design. The remaining ``bad'' pixels are classified as either hot, cold or non-linear, and are scattered relatively uniformly over the array, but with slight higher concentrations in the top right corner and a small region near the central of the array. 

As noted above, a typical pixel on the GPI detector has a dark current of less than 0.02 electrons per second. A pixel is considered ``hot" when it exhibits a highly elevated dark current. Maps of hot (high dark current) pixels are generated in the DRP using the ``Generate Hot Bad Pixel Map from Darks" recipe. The input files should be a set of 10 or more identical long ($t>100$ s) dark exposures. The algorithm measures the read noise from the standard deviation between frames and then locates pixels with high count rates significantly above this level. Specifically, the criteria for being considered a hot pixel is having a dark current greater than 1 electron per second, measured with greater than 5$\sigma$ confidence. A normal dark frame has $\sim$15,000 hot pixels. 

A pixel is classified as being ``cold", if its photo-sentivitity (response) is significantly below the average. Maps of cold (non-photo-sensitive) pixels are generated using the recipe ``Generate Cold Bad Pixel Map from Flats." Finding such pixels for GPI is more complicated than for typical instruments: Because of the fixed lenslet array, it is not possible to illuminate the detector evenly with any kind of flat illumination pattern. Regardless of the input, the light will always be focused by the microlenses creating a spatially variable illumination pattern with alternating illuminated and dark regions. However, by adding together many flat field exposures taken using several \textit{different} filters it is possible to illuminate (albeit non-simultaneously) all of the pixels on the detector. This illumination pattern is very structured and not a uniform illumination, therefore we refer to this as a ``multi-filter pseudo-flat". We then take advantage of the translational symmetries inherent in the lenslet array to build up a reference image that retains the spectral structure from the illumination pattern but is smoothed over several detector pixels. By comparing individual pixels to this reference image, we can identify those that lack sensitivity. The selection criterion to flag cold pixels is any pixel having a less than 15\% normalized response, as measured from the multi-filter pseudo-flat. The GPI detector has $\sim$2500 cold pixels.

A third class of bad pixels is those which show strongly nonlinear behavior. Of course, every pixel shows some non-linearity (see section \ref{sec:non_linearity}), however the pixels we are referring to here exhibit very significant non-linear behaviour at any level of exposure (without being strictly hot or cold). The software to measure these is not yet incorporated into the pipeline but exists as standalone code running outside of the pipeline. The non-linear pixels can be detected using a series of flat-fields taken while the detector was in ``write-all" mode. This engineering mode requires special intervention to enable but then causes each individual detector read in up-the-ramp mode to be written to disk. Examining the pixel response in each individual read allows identification of the non-linear pixels. This non-linear bad-pixel map is currently being supplied as a pre-created calibration file by the GPI instrument team to other users. The number of non-linear pixels is $\sim$1,000 and is not a significant contribution to the total number of bad-pixels.

An overall bad pixel map is generated by combining all three bad pixel types into a single boolean (1 or 0) image. During science reductions, the DRP can handle bad pixels by interpolating their values from surrounding pixels, using the primitive, ``Interpolate bad pixels in 2D frame." The method for interpolating is dependent upon the observing mode. For polarimetry mode, where each microlens forms a single Gaussian-like PSF, the interpolation is performed using all 8 surrounding pixels. In the case of spectral mode, each microlens forms a single long illumination pattern, so the interpolation utilizes only the vertical pixels. Certain pipeline tasks instead simply weight the bad pixels to zero, for instance the forward modeling wavelength calibration code \cite{Wolffthis}, and the functions that derive and fit the high-resolution microlens PSF models\cite{Ingraham2this}. The pipeline records which pixels have been interpolated in the data quality (DQ) FITS extension to make this record available to subsequent processing tasks.

Future work includes implementing as part of the pipeline the code for detection of the non-linear pixels, as well as a correction for the interpixel capacitance, as is done for the Wide Field Camera 3 on the Hubble Space Telescope \cite{Hilbert11a}.

\section{Read Noise}

\begin{figure}[ht]
\begin{center}
\begin{tabular}{cc}
\includegraphics[height=5cm, trim=0cm 0cm 0cm 0cm]{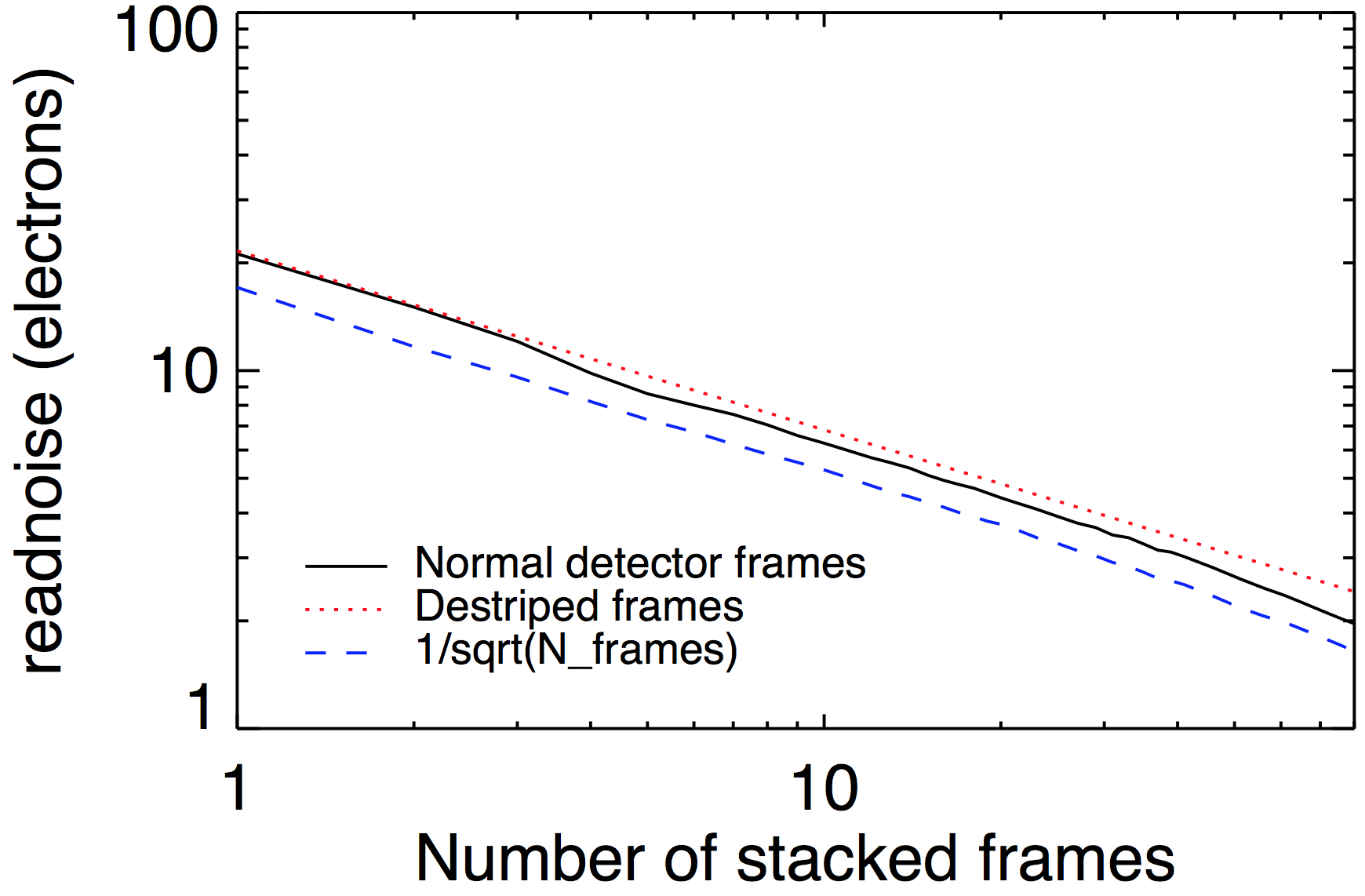} & \includegraphics[height=5cm, trim=0cm 0cm 0cm 0cm]{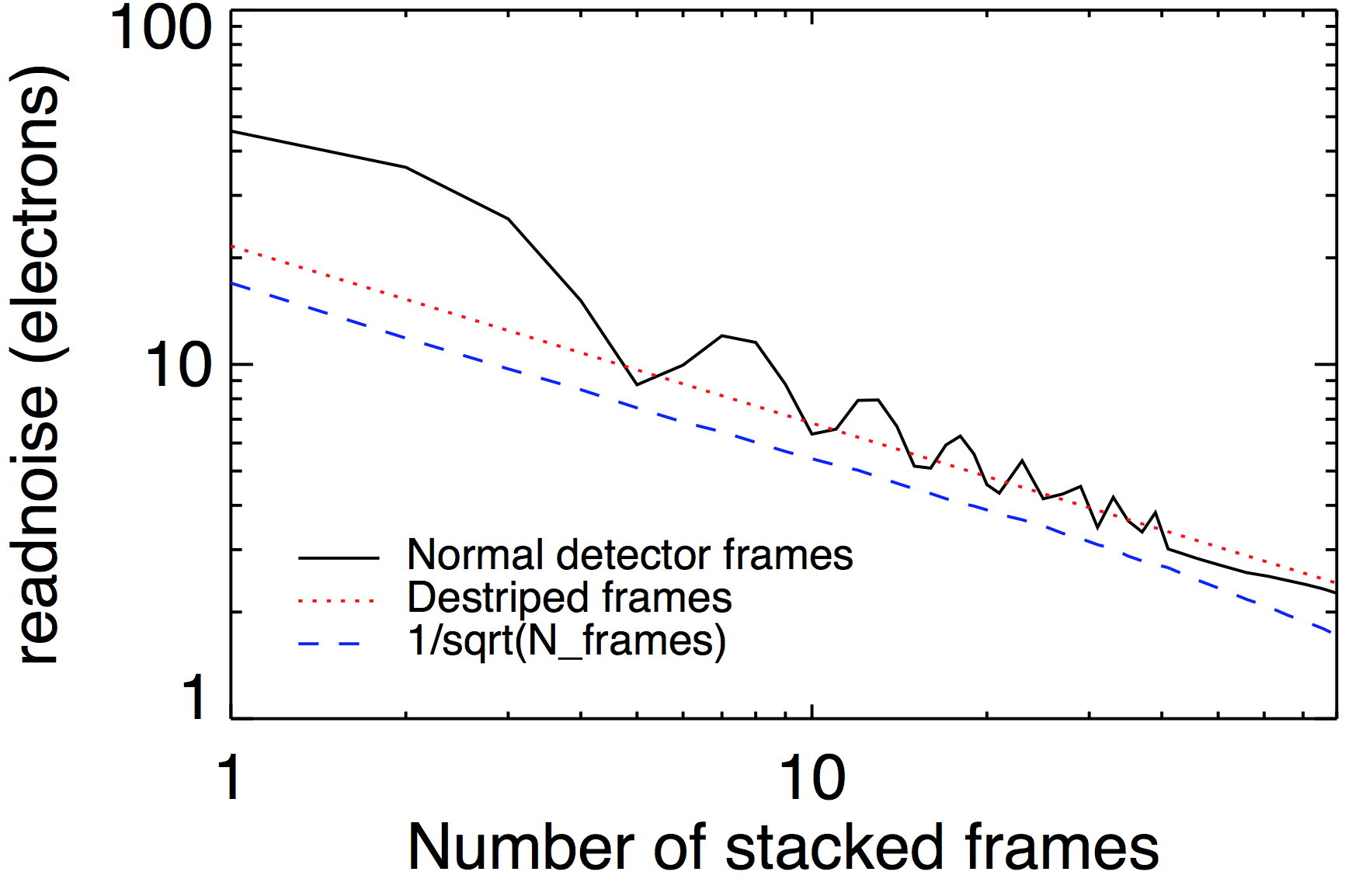} 
\end{tabular}
 
\end{center}
\caption[example]{ \label{fig:readnoise}  \textbf{Left:} The readnoise as a function of the number of averaged images in a region unaffected by microphonics. \textbf{Right:} The  same analysis in a region of strong microphonics. The black line indicates the noise in the raw detector images, whereas the blue dashed lines indicate the measurement in destriped images. The dotted red line indicates the expected $1/\sqrt{N_{frames}}$ behavior.}
\end{figure}

The detector readnoise for a single 1.5 s full-frame CDS exposure as produced by the detector electronics is 22 electrons RMS, with bad pixels given zero weight. After running the destriping algorithm, the readnoise is decreased to 17 electrons RMS. The variation over the field is small, with the exception of the areas that include the microphonics. To test how the readnoise averages down with multiple exposures, 250-1.5 s exposures were taken and averaged to make a single 1.5 s dark frame. This dark frame was subtracted from each exposure. The readnoise was then measured as a function of the number of images used in the average in a typical noise region and a microphonics noise region, both before and after destriping. Figure \ref{fig:readnoise} shows the readnoise behaviour over time for both sections. The black solid lines indicate the readnoise for the raw detector images, whereas the blue dashed lines indicate measurements on the destriped images. The dotted red line is a $1/\sqrt{N_{frames}}$ falloff with the intercept set to 22 electrons. 

The noise properties of the detector and readout electronics are very well behaved and average down as expected. The destriping routine does not impact this behaviour and therefore provides a significant improvement.

\section{Linearity}
\label{sec:non_linearity}

\begin{figure}[ht]
\begin{center}
\begin{tabular}{cc}
\includegraphics[height=7cm, trim=0cm 0cm 0cm 0cm]{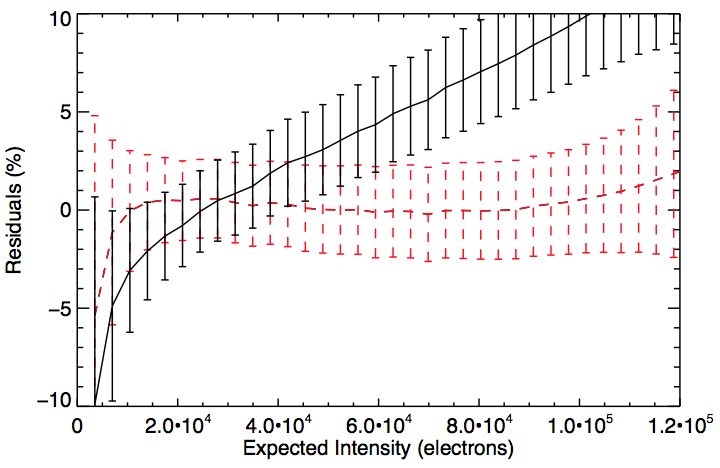}
\end{tabular}
\end{center}
\caption[example]{ \label{fig:nonlinearity} The non-linear behaviour as a function of incident intensity. The y-axis shows the percent error of the measured flux, where the black curve indicates the error when no non-linearity is applied. The red curve shows the percent error after a correction for non-linearity has been applied, which utilizes a simple quadratic response function.}
\end{figure}

The non-linear response for GPI’s H2RG, as it approaches full-well, (116,000 electrons according to the vendor report), was measured as part of early detector characterization studies\footnote{This study was performed while the detector electronics were tuned to mildly different settings than what is used today, therefore prior to implementing any of the corrections described below, a re-measurement will be performed.}. Upon putting the detector in it's ``write-all'' mode,  individual reads of an up-the-ramp sequence were stored and used in the analysis. The following measurements of the linearity were performed on the central 500$\times$500 pixel section of the array, using only the brightest pixel from each individual microspectrum. The H2RG is intrinsically non-linear due to the source follower architecture, but non-linearity is evident only near saturation, therefore, the intensity of the lamp was determined by fitting a slope to the measured intensity as a function of time for count levels between 2,000 and 10,000 ADU, where the detector response is expected to be very close to linear (the detector gain is 3.04 electrons/ADU). The very low count levels shown in figure \ref{fig:nonlinearity} are believed to be contaminated by persistence of previous images and detector noise artifacts, therefore they are not included in the analysis. The residuals\footnote{The plotted residuals are $\frac{\text{Incident intensity} - \text{Detected intensity}}{\text{Incident intensity}} \times 100$} of the incident intensity, based on the measurement of the source flux minus the measured intensity, are shown as the black curve in Figure \ref{fig:nonlinearity}. The analysis indicates that the detected flux exhibits an $\sim$3\% error at half-well ($\sim$19,000 ADU). This motivated all GPI observations thus far to keep the total count level for an individual frame to be below $\sim$16,000 ADU. It should be noted that this count level is only reached during science observations in polarimetry mode, or for bright targets in spectral mode. In general, the field rotation dictates the maximum exposure time. If longer exposures are possible without blending due to field rotation, then it has been thus far recommended to use coadditions to minimize observing overheads.

In order to explore the possibility of exposing to greater filled well depths, a correction was derived to compensate for the non-linearity. A simple quadratic function ($ADU' = a_0 + a_1 * ADU + a_2 * ADU^2$) was used to model the non-linearity over the entire the 2,000-28,000 ADU range. When this correction is applied to the data, the non-linearity is very well removed to $\sim$80,000 electrons, as shown by the red curve in figure \ref{fig:nonlinearity}. This indicates that with a properly calibrated correction, exposing to greater count levels is possible, and may benefit the spectral observations of bright targets, or observations in polarimetric mode.

\section{Persistence}

Persistence in HgCdTe arrays is a well documented phenomena. Persistence is a particular risk for high-contrast instruments such as GPI, where we might be imaging a star followed immediately by a planet 15 magnitudes fainter. Operational procedures (e.g. use of shutters to avoid illumination by a star before it is centered behind the coronagraph mask) mitigate this, but it is still an issue.
The amount of persistence and whether or not a correction is necessary depends upon the illumination history of the detector and the exposure times of the science images of interest. The physics behind the effect of persistence is complex and still not fully understood, however, a detailed qualitative model exists that offers an accurate description of the effect \cite{Smith08a}. In short, persistence results from charges trapped in the depleted negative portion of the PN junction of the array. The amount of charge captured depends on the charge level accumulated on the pixel and the amount of time that charge is present. The amount of persistence is observed to vary between pixels and various arrays. When the detector is reset, the trapped charge begins to release, acting as an increased dark current decaying slowly over time. Following cases of  highly saturated exposures, persistence on the GPI detector has been observed well above the readnoise level for over 10 minutes after the illumination was removed.

Preliminary analysis of the properties of the GPI detector has demonstrated that the persistence is very low and in many cases correction for this effect is not needed. However, obtaining the highest precision spectrophotometry of bright targets with GPI will most likely require a persistence correction. Fortunately, deriving a correction for the persistence does not rely on a full physical understanding of the persistence mechanism. Using the observed properties, it is possible to obtain a numerical model predicting its behaviour over time. For GPI, we follow the model derived for the Hubble Space Telescope’s Wide Field Camera 3 \cite{Rajan11a}\cite{Long13a}. The parameters for the model were derived based on illuminating the detector to its half-well capacity, then examining a series of frames every 8 seconds to measure the decline in the persistence. 
Although the persistence depends upon the stimulus, the rate of decay appears relatively independent of stimulus and is observed to reduce by a factor of 2 approximately every 2 minutes. 

\begin{figure}[ht]
\begin{center}
\begin{tabular}{cc}
\includegraphics[height=6cm, trim=0cm 0cm 0cm 0cm]{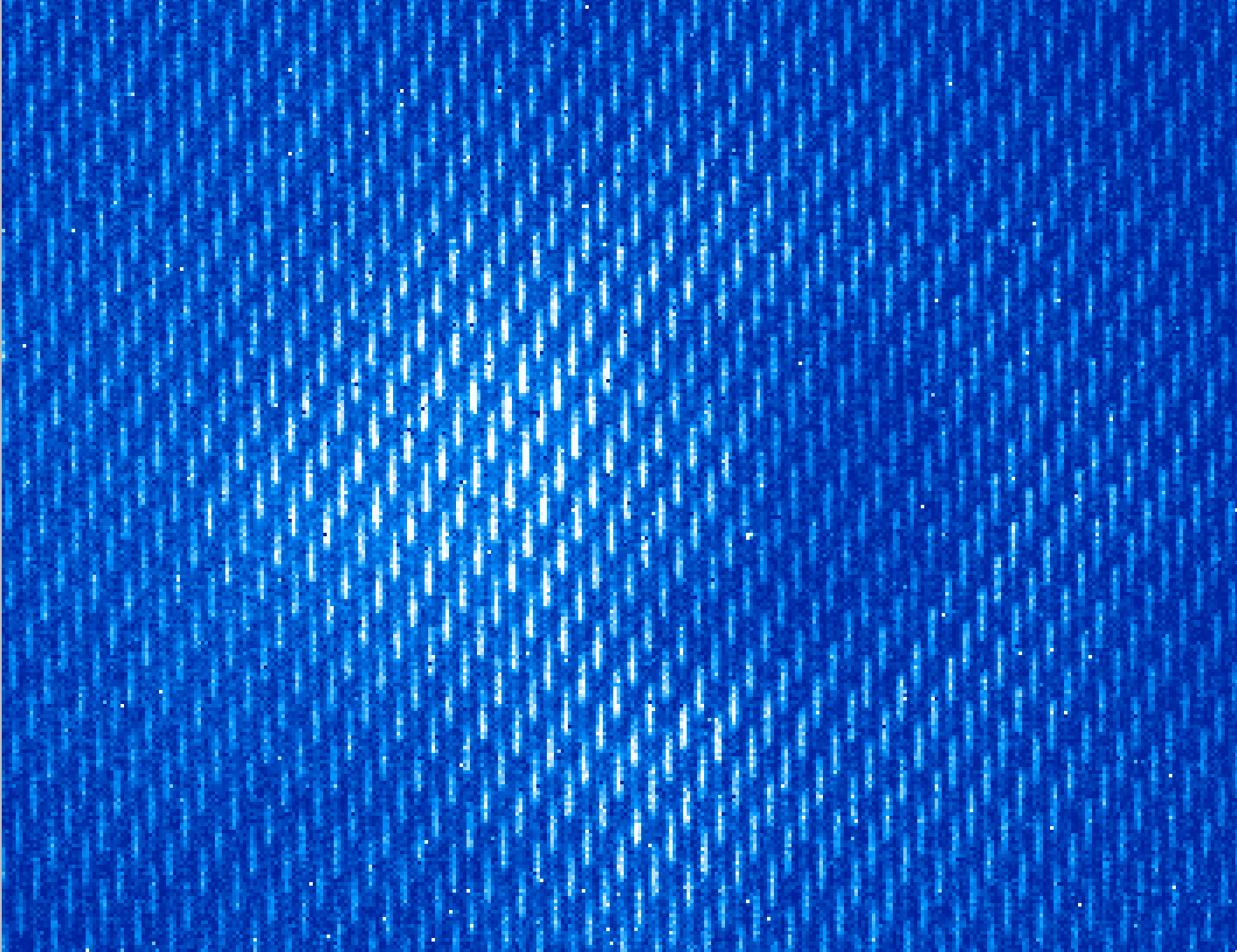} & \includegraphics[height=6cm, trim=0cm 0cm 0cm 0cm]{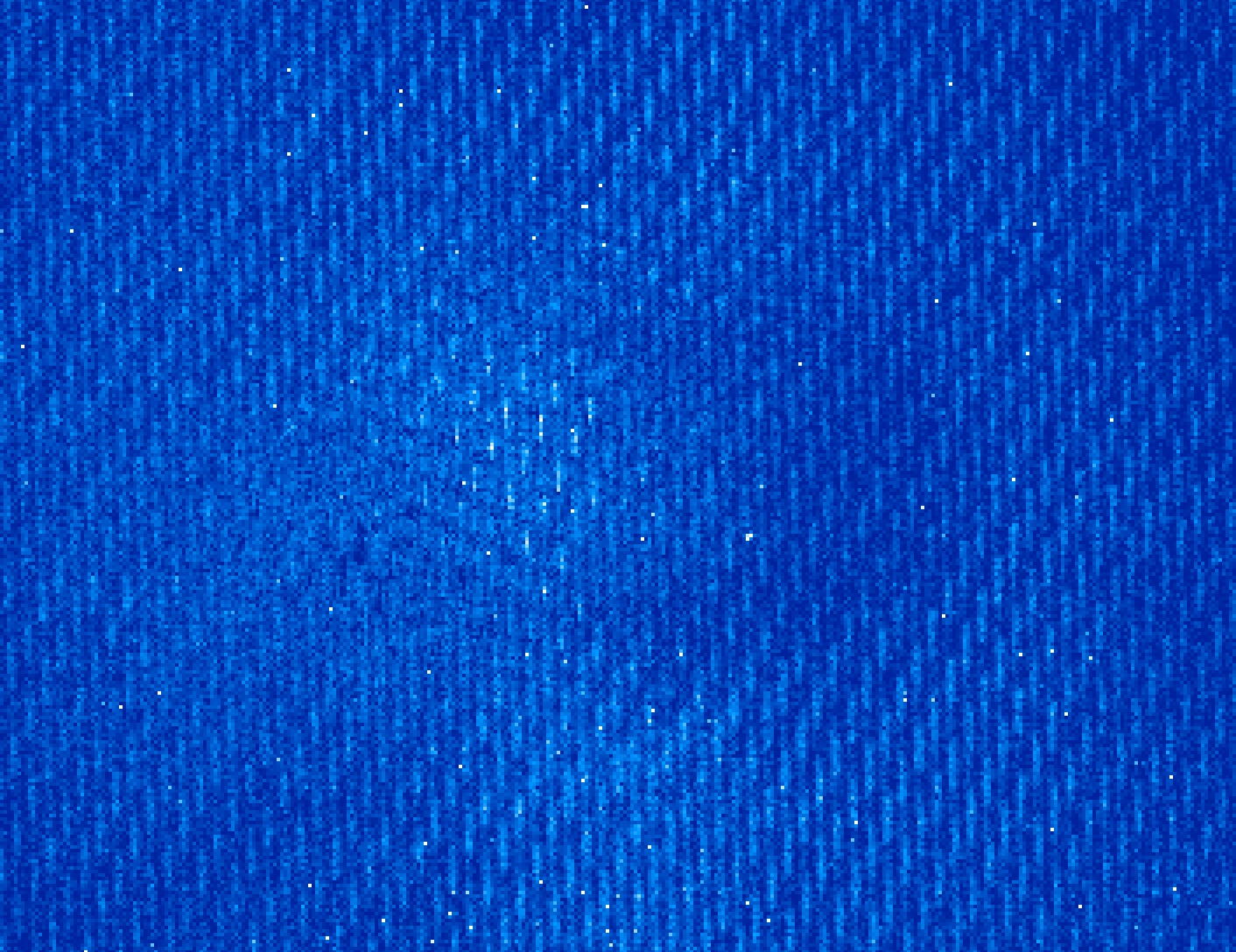}
\end{tabular}
\end{center}
\caption[example]{ \label{fig:persistence} \textbf{Left:} The central portion of a 60 second (42 up-the-ramp read) $Y$-band sky frame showing persistence from the previous coronagraphic image taken 5 minutes before.  \textbf{Right:} The same sky image after correction for the persistence using the GPI DRP.}
\end{figure}

The pipeline primitive, ``Remove Persistence from Previous Images,'' uses this model to estimate and subtract persistence in any given science image, given the series of exposures preceeding that image. Figure \ref{fig:persistence} shows current effectiveness of the model to remove the persistence. 

Current performance of the persistence removal is limited by the data used to derive the model. The 8 second sampling does not provide adequate coverage to enable precise measurements of the exponential falloff. Because data was only taken in a single band, a large portion of the array was never illuminated, and therefore we were forced to spatially bin the model to 64x64 pixel boxes. This results in an over- or under-subtraction in many parts of the array where the persistence properties of the detector change on small spatial scales. Another limitation is that users will never have a full record of the illumination history of the array, since the array is not read out after every detector reset.

Future work will extend persistence calibration to the entire array and further explore the amount of persistence as a function of varying fluence, or stimulus. Lastly, we will obtain test data with increased time sampling to better sample the exponential decay. 

\section{Discussion}

While the HAWAII-2RG detector is state of the art, it is still subject to many imperfections and systematics. For GPI, depending on the details and science goals of any particular observations, some of these systematics may be ignored without adverse impact but in other situations they must be attended to.  For instance, the impact of a handful of uncorrected bad pixels is generally very minor for a long exposure ADI sequence as part of a planet search: the PSF subtraction naturally subtracts out any residual that is fixed in the detector frame, while the signal from a planet will be spectrally dispersed across many pixels and will move across even more different detector pixels as the sky rotates. In contrast for observations of a resolved solar system object, uncorrected bad pixels directly reduce cosmetic quality and S/N for measurement of surface features, and no PSF subtraction is performed.
In polarization mode, the non-common-path offset between the orthogonal polarization channels is dominated by detector artifacts such as cold pixels that affect one polarization but not the other. Users of GPI should apply their own scientific judgement when determining which correction steps are needed in their particular reduction recipes.

Some aspects of detector calibration for GPI are relatively mature, simply needing ongoing monitoring for the long term, for instance the routine acquisition of darks. Other aspects have more substantial analysis work yet to be done, for instance the planned improvements to persistence calibration mentioned above, and better nonlinearity correction. We currently neglect the detection and correction of cosmic rays. The observed rate of cosmic ray induced electrons is quite low for GPI, given its substrate removed detector, but it is not zero and occasionally cosmic rays are visible particularly in longer exposures.  Interpixel capacitance is not accounted for, nor intrapixel QE variations (with the exception that these latter are implicitly included when deriving high resolution microlens PSF models, although not explicitly considered in any reduction step).  In some cases we plan to adapt algorithms developed for HST and JWST detector calibration for use with GPI data.

A particular challenge is detector flat fielding. Detector flat fields cannot be directly measured now; as noted above given the lenslet array there is no way to evenly illuminate the detector. The multi-filter pseudo-flat technique described above works to locate grossly low QE pixels but is not precise enough to measure smaller QE variations between good pixels.  Given the presence of flexure continuously changing the registration between lenslets and the detector it is not sufficient to take day calibration flats while the telescope is parked at zenith, since slightly different sets of pixels will be illuminated at night. Improved calibration methods to deal with this situation are under investigation.

\section{Acknowledgments}
We would like to thank the staff of the Gemini Observatory for their assistance in the characterization of the detector. The Gemini Observatory is operated by the Association of Universities for Research in
Astronomy, Inc., under a cooperative agreement with the NSF on behalf of the Gemini
partnership: the National Science Foundation (United States), the National Research
Council (Canada), CONICYT (Chile), the Australian Research Council (Australia),
Minist\'erio da Ci\'encia, Tecnologia e Inova\c{c}\=ao (Brazil), and Ministerio de Ciencia,
Tecnolog\'ia e Innovaci\'on Productiva (Argentina).

\bibliography{main}   
\bibliographystyle{spiebib}   

\end{document}